\documentclass[aps,prl,preprint,superscriptaddress]{revtex4}

\usepackage{graphicx}
\usepackage{dcolumn}
\usepackage{bm}
\usepackage{amsmath}
\usepackage{amssymb}
\usepackage{latexsym}
\usepackage{epsfig}
\usepackage{amsbsy}
\usepackage{array}
\usepackage{amssymb}
\usepackage{setspace}
\usepackage{bm}

\def\sint{\ifmmode{- \!\!\!\!\!\! \int}
    \else{\hbox{$- \!\!\!\! \int \ $}}\fi}

\begin{document}

\title{Bounding the speed of `spooky action at a distance'}

\author{Juan Yin, Yuan Cao, Hai-Lin Yong, Ji-Gang Ren, Hao Liang, \\Sheng-Kai
Liao, Fei Zhou, Chang Liu, Yu-Ping Wu, Ge-Sheng Pan, Li Li, \\Nai-Le Liu$^{*}$, Qiang Zhang$^{*}$, Cheng-Zhi Peng$^{*}$ and Jian-Wei Pan}
\renewcommand{\thefootnote}{\fnsymbol{footnote}}
\footnotetext[1]{Author to whom correspondence should be addressed; electronic mail: pcz@ustc.edu.cn, qiangzh@ustc.edu.cn and nlliu@ustc.edu.cn.}

\affiliation{Shanghai Branch, National Laboratory for Physical Sciences at Microscale, and Department of Modern Physics, University of Science and Technology of China, Shanghai 201315, China}


\begin{abstract}
In the well-known EPR paper, Einstein et al. called the nonlocal correlation in quantum entanglement as `spooky action at a distance'. If the spooky action does exist, what is its speed? All previous experiments along this direction have locality and freedom-of-choice loopholes. Here, we strictly closed the loopholes by observing a 12-hour continuous violation of Bell inequality and concluded that the lower bound speed of `spooky action' was four orders of magnitude of the speed of light if the Earth's speed in any inertial reference frame was less than $10^{-3}$ times of the speed of light.
\end{abstract}



\maketitle


In order to test the speed of `spooky action at a distance'\cite {EPR35}, Eberhard proposed\cite{Eberhard89} a 12-hour continuous space-like Bell inequality \cite{Bell64,CHSH} measurement over a long east-west oriented distance. Benefited from the Earth self rotation, the measurement would be ergodic over all possible translation frames and as a result, the bound of the speed would be universal\cite{Eberhard89,Salart08} .

Salart et al.\cite{Salart08} recently report to have achieved the lower bound of `spooky action' through an experiment using Eberhard's proposal. In that experiment, time-bin entangled photon pairs were distributed over two sites, both equipped with a Franson-type interferometer\cite{Frason89} to analyze the photon pairs. At one site, the phase of the interferometer was kept stable and the phase of the interferometer at the other site was continuously scanned. An interference fringe of 87.6\% was achieved, which was high enough for a violation of the CHSH-Bell inequality. However any bipartite Bell test requires at least two settings at each site because if there was only one setting, one could always achieve the same interference fringe with a product state without any entanglement. Moreover, as pointed out by Kofler et al.\cite{Comment08}, even if the phase of the first site had a second setting, the experiment still had locality loophole, because the setting choice on one site could not be space-like separated from either the measurement events at the other site or the entanglement source generation event\cite{Clauser74,Bell85,Scheidl2010}. Due to these loopholes, the experiment could be explained by common causes\cite{Comment08} instead of `spook action'. Actually, similar loopholes existed in all previous attempts for the speed measurement of `spooky action'\cite{Gisin2000,Gisin00,Cocciaroa}.


Here, we distributed entangled photon pairs over two site that were 16-km apart sites via free space optical link and implemented a space-like Bell test \cite{Bell64,CHSH} to close the locality loophole. Meanwhile, we utilized fast electro optic modulators (EOMs) to address the freedom-of-choice loophole. As almost all the photonic Bell experiment\cite{Scheidl2010,Aspect82, Weihs98,Clauser78}, we utilized fair sampling assumption to address the detection loophole\cite{Reply08}. This assumption is justified by J. S. Bell himself's comments\cite{Bell87,Weihs98}, ``\ldots it is hard for me to believe that quantum mechanics works so nicely for inefficient practical set-ups and is yet going to fail badly when sufficient refinements are made. Of more importance, in my opinion, is the complete absence of the vital time factor in existing experiments. The analyzers are not rotated during the flight of the particles.''

Our two sites are east-west oriented sites at the same latitude as Eberhard proposed\cite{Eberhard89}. Suppose two events occur in the two sites, respectively at positions $\vec{\bm r_A}$ and $\vec{\bm r_B}$ at time t$_A$ and t$_B$ under an inertial reference frame centered on the Earth and given another inertial reference frame L, moving with a speed $-\vec{\bm\upsilon}$ relatively to the Earth center frame, the two events' coordinates are ($\vec{\bm r_A}^\prime,t_A^\prime$) and ($\vec{\bm r_B}^\prime,t_B^\prime$). When Bell inequality violation is observed with the two events in frame L, the speed of `spooky action' $\bm {V_{sa}}$ has a lower bound shown as below:
\begin{equation}
\bm {V_{sa}} \geq \frac {\parallel \vec{\bm r_B}^\prime-\vec{\bm r_A}^\prime \parallel} {|t_B^\prime-t_A^\prime|}
\end{equation}

Under the Earth-centered inertial coordinate frame, the lower bound is obtained by Lorentz transformation as follow:
\begin{equation}
(\frac {V_{sa}} {c})^2 \geq 1+\frac {(1-\beta^2)(1-\rho^2)} {(\rho+|\beta_{//}|)^2}
\end{equation}

where, c is the speed of light; $\rho=c\cdot |t_B-t_A|/\parallel \vec{\bm r_B}^\prime-\vec{\bm r_A}^\prime \parallel$ is a factor of alignment of the two events in Earth center frame; $\beta=|\vec{\bm\upsilon}|/c$ is the Earth center's relative speed in the L frame and $\beta_{//}=|\vec{\bm\upsilon}_{//}|/c$ with $\vec{\bm\upsilon}_{//}$ the component of $\vec{\bm\upsilon}$ parallel to the A-B axis. Due to the Earth self rotation, $\beta$ and $\beta_{//}$ are time dependent. Suppose the angle between A-B axis and Earth center frame's X-Y axis is $\alpha$, the angle between $\vec{\bm\upsilon}$ and Earth center frame's Z axis is $\theta$ and $\omega$ is the angular velocity of the Earth, we have:
\begin{equation}
\beta_{//}(t)=\beta\cos\theta\sin\alpha + \beta\sin\theta\cos\alpha\cos\omega t
\end{equation}

When A-B axis is exactly east-west oriented, $\alpha=0$ and equation (3) is simplified into:
\begin{equation}
\beta_{//}(t)=\beta\sin\theta\cos\omega t
\end{equation}

If we choose the measuring period T to meet $\omega T \ll 1$, there always has a time $t$ in the period to satisfy $|\beta_{//}| \leq |\beta \sin\theta \cdot \omega T/2|$. Therefore we achieve the optimal lower bound for $\bm {V_{sa}}$, which is:
\begin{equation}
(\frac {V_{sa}} {c})^2 \geq 1+\frac {(1-\beta^2)(1-\rho^2)} {(\rho+|\beta \sin\theta \cdot \omega T/2|)^2}
\end{equation}

Fig. 1 was the setup of the whole experiment. We, at the sending site, prepared and distributed entangled photon pairs to two receiving sites through two refractor telescopes respectively. As was shown in Fig. 1a., all the sites were located at the Qinghai Lake region, Qinghai Province, P. R. China. The two receiving site were east-west oriented and separated with a 15.3 km distance. One was a scenery view point (36$^{\circ}$33$^{\prime}$15.31$^{\prime\prime}$N, 100$^{\circ}$28$^{\prime}$24.66$^{\prime\prime}$E) and the other was a fishing farm (36$^{\circ}$33$^{\prime}$15.31$^{\prime\prime}$N, 100$^{\circ}$38$^{\prime}$42.00$^{\prime\prime}$E). The sending site(36$^{\circ}$32$^{\prime}$25.22$^{\prime\prime}$N, 100$^{\circ}$33$^{\prime}$33.30$^{\prime\prime}$E) was in the middle of the two receiving sites.

At the sending site, we generated polarization entangled photon pairs via type II spontaneous parametric down conversion (SPDC)\cite{Kwiat95} in a periodically poled KTiOPO$_4$ (PPKTP) crystal. In Fig. 1b., a single-longitudinal-mode ultraviolet semiconductor laser (405nm) was 45$^\circ$ linearly polarized by a half wave plate (HWP) and a quarter wave plate (QWP) and then splited into two beams by a polarization beam splitter (PBS). After passing through the PPKTP crystal placed in the center of the Sagnac interferometer\cite{Fedrizzi07}, the photon pairs generated by SPDC from opposite directions interfered at the same PBS and formed the polarization entanglement. The photon pairs were in the two-particle singlet state, i.e. $|\psi^{-}\rangle=(|H\rangle_A|V\rangle_B-|V\rangle_A|H\rangle_B)/\sqrt 2$. Under a 12.4 mW pump power, we achieved 550 KHz coincidence rates with 3-nanosecond coincidence window. The visibility was better than 91.3\%(accidental counts not subtracted). The PPKTP crystal was set at its degenerate temperature by a thermoelectric cooling (TEC) temperature controller and both converted photon pairs were at the degenerate wavelength 811.54 nm. The full width at half maximum (FWHM) of the photon pairs' spectral linewidth was 0.6 nm , which was decided by phase-matching condition of the SPDC process. Finally, the entangled photon pairs were coupled into single mode fibers (SMFs) and led to the integrated sending system, which consisted of an 80-mm f/7(that is, aperture 80mm, focal length 560mm) extra-low dispersion apochromatic refractor telescope, a dichroic mirror (DM) and a piezo-driven fast steering mirror (FSM). The divergence of our compact quantum transmitter was about  30~$\mu$rad.

The distributed entangled photons were collected by receiving telescopes and delivered to polarization analyzing units at both receiving sites. In each of the unit (Fig. 1c), a random number controlled EOM combined with a PBS was applied to actively choose the measuring base of the Bell inequality. The outputs of the PBS were coupled into two 105~$\mu$m core diameter, multi-mode fibers through aspherical lens, and thereafter detected by the single-photon counting modules (SPCMs), respectively. The detection signals from SPCMs were encoded with the controlled signals applied in EOMs as the output ones.

As was described before, in order to be ergodic all possible inertial reference frame, a continuous 12-hour Bell inequality measurement was necessary. However, the shaking of the whole system and the atmosphere turbulence influenced the telescope optical link. In previous experiments\cite{Peng05,Aspelmeyer03b}, the free space optical link would lose after at most ten-minute continuous measurement, which was definitely not long enough for our experiment. Therefore, we developed laser assisted free space tracking techniques as shown in Fig. 1a. We sent a green lasers light (532 nm, 100 mw, 1.5 mrad) from sending site to the two receiving sites, respectively as the tracking signal. Meanwhile, in both receiving sites, we sent a red laser light (671 nm, 100 mw, 1.5 mrad) back to the sending site. All lasers were sent co-axially with the entangled photon pairs. In each sending or receiving site, we utilized a Complementary Metal-Oxide-Semiconductor (CMOS) camera to detect the 532 nm laser signal and a two-dimensional rotatable platform in both azimuth and elevation to track, which constituted a closed loop coarse tracking system.

The bandwidth of coarse tracking system was about 10 Hz, which was decided by the read-out time of the camera. The slow response time of the system could not keep the free space optical link if some sudden change, for example the atmosphere turbulence, happened in the channel. In order to increase the tracking system's bandwidth, we adopted a fine tracking system in each sending telescope. The fine tracking system was composed of a four-quadrant detector(QD), a piezo-driven FSM and a DM. The tracking light (671 nm) from the receiving site, was collected by the sending telescope and then reflected by the FSM into the DM. The DM transmitted the entangled photons and reflected the tracking light into the QD. The QD's detection information would feedback to the FSM to make a close loop. Then thus, we achieved the tracking accuracy better than 3.5~$\mu$rad with more than 150 Hz bandwidth and the continuous measurement time extended beyond 12 hours.

In order to satisfy the space-like criteria, we needed to strictly time tag all the events during the measurement. In the experiment, we utilized laser synchronization technology. In the sending site, we split and sent co-axially with the entangled photons, a near infrared pulse laser (1064 nm, 5 kHz, 10 mw, 200~$\mu$rad) to two receiving sites as the synchronization signal. The signals were detected by high-speed avalanche photo diodes. Meanwhile, both receiving sites had GPS (Global Position System) for the system's initial synchronization and rough timing adjustment. Both sites utilized time to digital converters (TDC) to record the detection signals, synchronization signals and GPS signals. Therefore, we achieved a timing accuracy better than 1 ns.

Fig. 2 was the space-time diagram of the experiment. We set the entangled photon pairs generation as the original point of the space-time diagram. The points A (-7.8 km, 26.1~$\mu$s) and B (7.8 km, 26.1~$\mu$s) represented the measurements in receiving site A and B, respectively. By adjusting the fiber length at the sending site, we distributed the entangled photon pairs to both receiving sites almost at the same time. The overall timing uncertainty between A and B was 350 ps (corresponding to a distance of 105mm), which was decided by the timing jitter of the detectors and TDCs. The time difference between A and B events was far less than their space distance. So the two measurement events A and B were space-like separate.

Points a (-7.8 km, 23.1~$\mu$s) and b (7.8km, 23.1~$\mu$s) represented the setting choice events in site A and B, respectively. The dotted green line D between b and B represented electronic delay from the quantum random number generator (QRNG) to EOM. The solid red line d  represented the total optical delay of sending and receiving system. According to the figure, D was much larger than d, so that the selection choice event was out of the light cone E and then thus the setting choice event would never be influenced by the generation of entangled source. Furthermore, event A was out of light cone b and event B was out of cone a, which meant that the selection choice a (b) would not influence the measurements in site B (A). Therefore we closed both locality and the freedom-of-choice loopholes in our experiment. As space-like separation was invariant under Lorentz transformation, the loopholes were closed in all inertial reference frames.

Then, we tested Bell-CHSH inequality, which was:
 \begin{equation}
S=|E(\theta_A,\theta_B)+E(\theta_A^{\prime},\theta_B)+E(\theta_A,\theta_B^{\prime})+E(\theta_A^{\prime},\theta_B^{\prime})|\leq 2
\end{equation}

Here, $\theta_A$,$\theta_A^{\prime}$($\theta_B$,$\theta_B^{\prime}$)were the polarization basis for the measurement setting in site A (B), and $E(\theta_A,\theta_B)$ was the expectation value of the correlation between polarization measurements A and B. According to quantum mechanics, when $\theta_A=0$, $\theta_A^{\prime}=\pi/4$, $\theta_B=\pi/8$, $\theta_B^{\prime}=-\pi/8$, $S=2\sqrt 2$ and a maximum violation of CHSH inequality could be achieved.

Fig. 3 showed 12-hour continuous experimental data for the CHSH inequality violation. When the Bell inequality violation was guaranteed, we could calculate the speed lower bound of `spooky action'. According to equation (5), the lower bound was related to $\theta$, the angle between A-B axis and Z axis of Earth center frame, and $\beta$, the Earth center's relative speed to any inertial reference frame. Fig. 4 showed different speed lower bound with different $\theta$ and $\beta$.

According to the space-time diagram, $\rho=c\cdot {350ps}/{15.35km}=6.84\times 10^{-6}$. When, $\beta$ was considered as 10$^{-3}$, similar to the Earth's center speed under the 2.7 K cosmic microwave background reference frame. Given $\theta=\pi/2$, we achieved the lower bound for $\bm {V_{sa}}$ via equation (5):
\begin{equation*}
\frac {V_{sa}} {c} \geq \frac {1} {\rho+|\beta| \cdot \omega T/2}=1.38\times 10^4
\end{equation*}

We could also conclude from Fig. 4, when the Earth center's relative speed reach 0.9c, the speed of spook action would still be 7 times higher than the speed of light.

Therefore, we experimentally achieved the lower bound of the `spooky action' speed by distributing polarization entangled photon pairs over two exactly east-west oriented sites and observed a 12-hour continuous space-like Bell inequality violation. It should be noted that, with only two parties one may always hold the opinion that a real fast enough "spooky action" could explain quantum correlations violating Bell inequality without leading to superluminal communication. Recently some researchers found that this no longer holds when more parties are involved\cite{Bancal12}.

The developed fast time tagging, long-distance laser tracking and synchronizing technology in the experiment can also find immediate applications in the satellite-earth quantum communication\cite{Aspelmeyer03b}, multi-party quantum communication\cite{Yuao05} and test of space-like GHZ theorem\cite{Bouwmeester99}.



We are grateful to the staff of the Qinghai Lake National Natural Reserve Utilization Administration Bureau, especially Y.-B. He and Z. Xing, for their support during the experiment. We acknowledge insightful discussions with Y.-A. Chen and B. Zhao. This work has been supported by the National Fundamental Research Program (under grant no. 2011CB921300 and 2013CB336800), the Chinese Academy of Sciences and the National Natural Science Foundation of China.



\newpage

A list of figures

\begin{enumerate}

\item Diagram of experimental setup for testing the speed of spooky action.
 (a) Bird's-eye view of the experiment. We generated entangled photon pairs in the sending point and utilized an integrated sending system to distribute them into two east-west oriented receiving points A and B. Receiving site A used an off-axis reflection telescope with 400mm diameter, and B used a refraction telescope with 127mm diameter. Both receiving sites A and B had polarization analyzing units (the orange box ).
 (b) Entanglement generation setup. A pair of 45-degree mirrors and a dual-wavelength PBS formed the Sagnac interferometer. A 22.5$^\circ$ dual-wavelength HWP was used to swap the horizontal and vertical polarization. The DM was used to distinguish the pump light and the signal light. The entangled photons were coupled to SMFs through aspherical lens.
 (c) Active polarization analyzing unit in receiving site A. A HWP, an EOM, a PBS, and two multi-mode fiber-coupled single photon counting modules (SPCMs) constitute an active polarization analyzing unit. A quantum random number generator (QRNG) and an amplifier were used to drive the EOM. The signals from the SPCMs combined the logic module were sent to a time digital converter (TDC).
 (d) Active polarization analyzing unit in receiving site B. \label{fig:Graph1}

\item Space-time diagrams of the speed measurment experiment. Points A and B indicated measurement events at sites A and B respectively. Points a and b indicated A and B\rq s setting choice. Point E represented the entanglement generation.\label{fig:Graph2}

\item 12-hour continuous experimental data for the CHSH inequality violation. Each blue data point represented the value of S in one time interval T, 1800 seconds in our experiment. The red line represented the classical limit. According to the picture, all 26 data points violated the Bell-CHSH inequality by 1$\sim$10 standard deviations above the local realistic bound of 2, depending on weather conditions. In order to reduce the day light noise, we implemented experiments from 19 pm to 7 am of the next morning.\label{fig:Graph3}

\item The speed of spooky action at different $\beta$ and $\theta$. For any $\beta$, the speed lower bound has a minimum when $\theta$ was $\pi/2$. And as long as $\beta<1$ , i.e. the Earth's center speed was less than the speed of light, the speed of Òspooky actionÓ was beyond the speed of light.\label{fig:Graph4}

\end{enumerate}

\newpage

\begin{figure}
\centering
\begin{minipage}[b]{0.5\textwidth}
\centering
\includegraphics[width=3in]{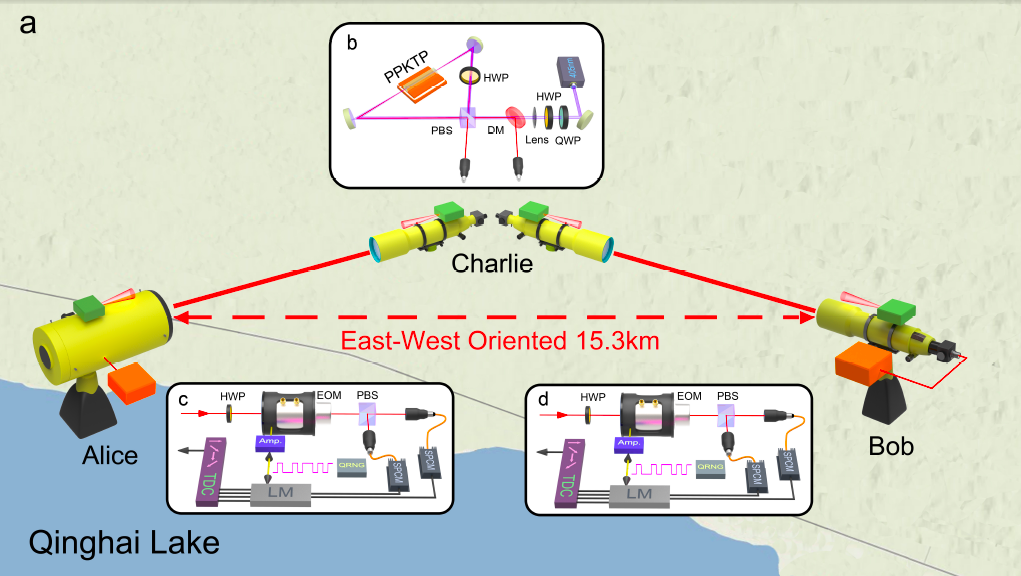}
\end{minipage}
\caption{Diagram of experimental setup for testing the speed of spooky action. \label{fig:Graph1}}
\end{figure}

\begin{figure}
\centering
\begin{minipage}[b]{0.5\textwidth}
\centering
\includegraphics[width=3in]{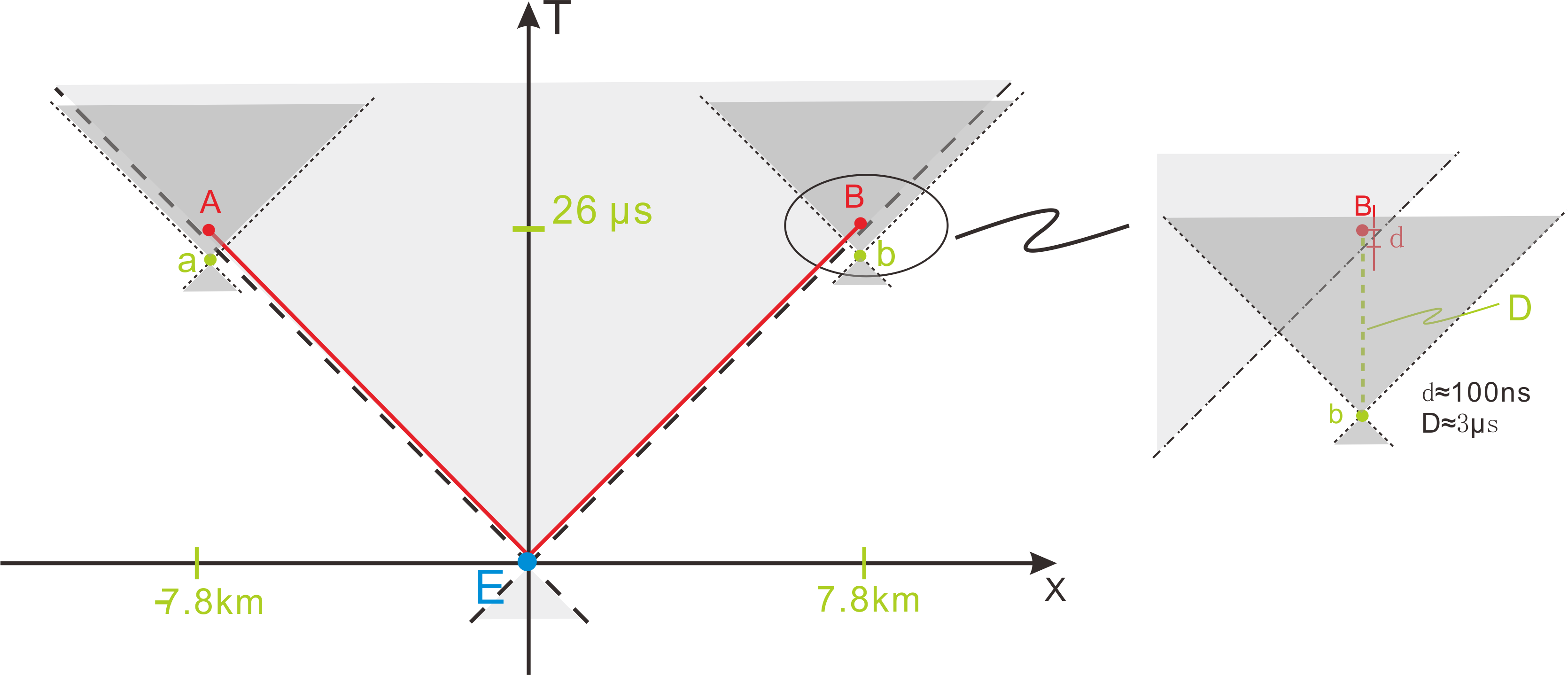}
\end{minipage}
\caption{Space-time diagrams of the speed measurment experiment. \label{fig:Graph2}}
\end{figure}

\begin{figure}
\centering
\begin{minipage}[b]{0.5\textwidth}
\centering
\includegraphics[width=3in]{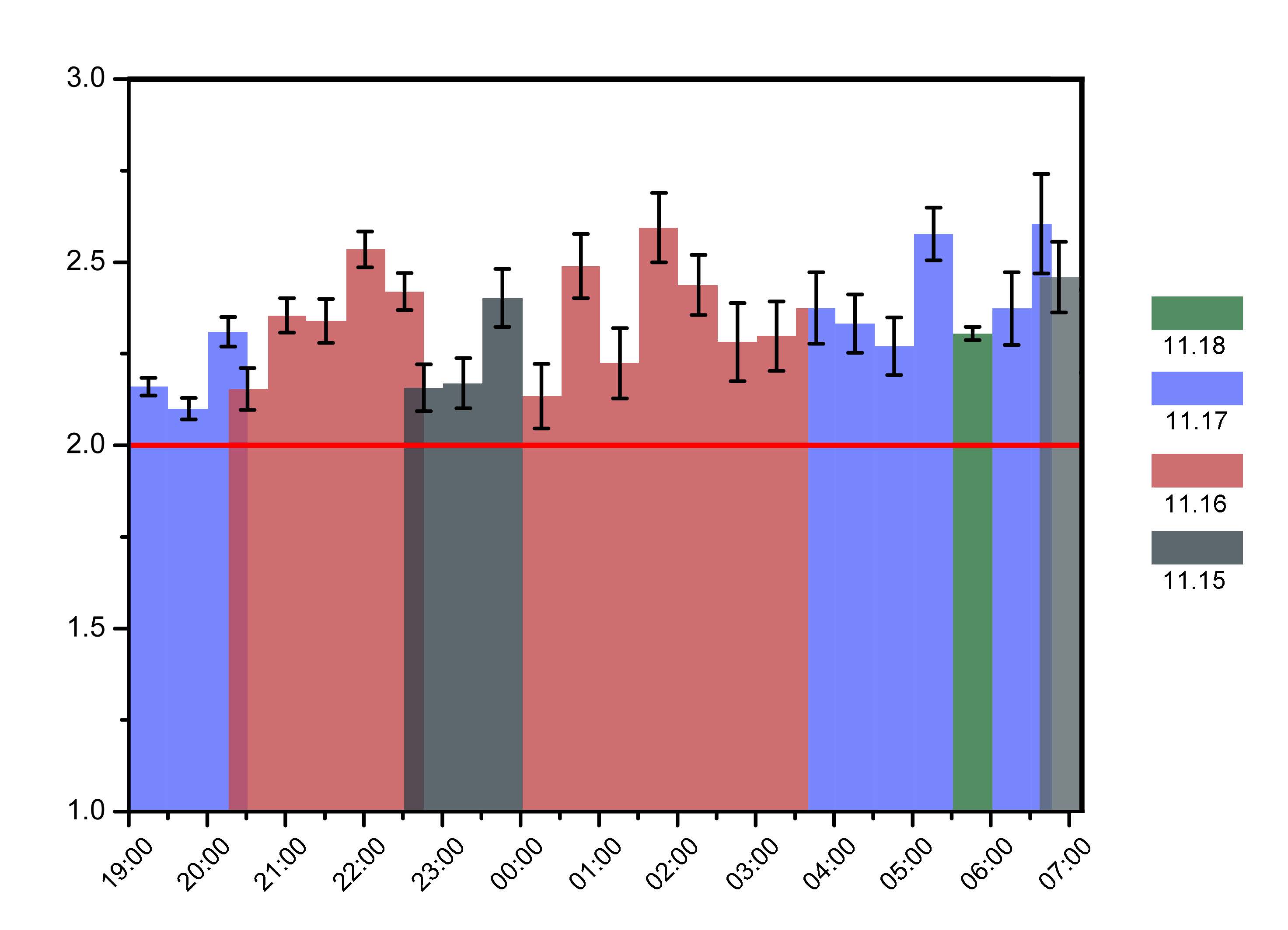}
\end{minipage}
\caption{12-hour continuous experimental data for the CHSH inequality violation. \label{fig:Graph3}}
\end{figure}

\begin{figure}
\centering
\begin{minipage}[b]{0.5\textwidth}
\centering
\includegraphics[width=3in]{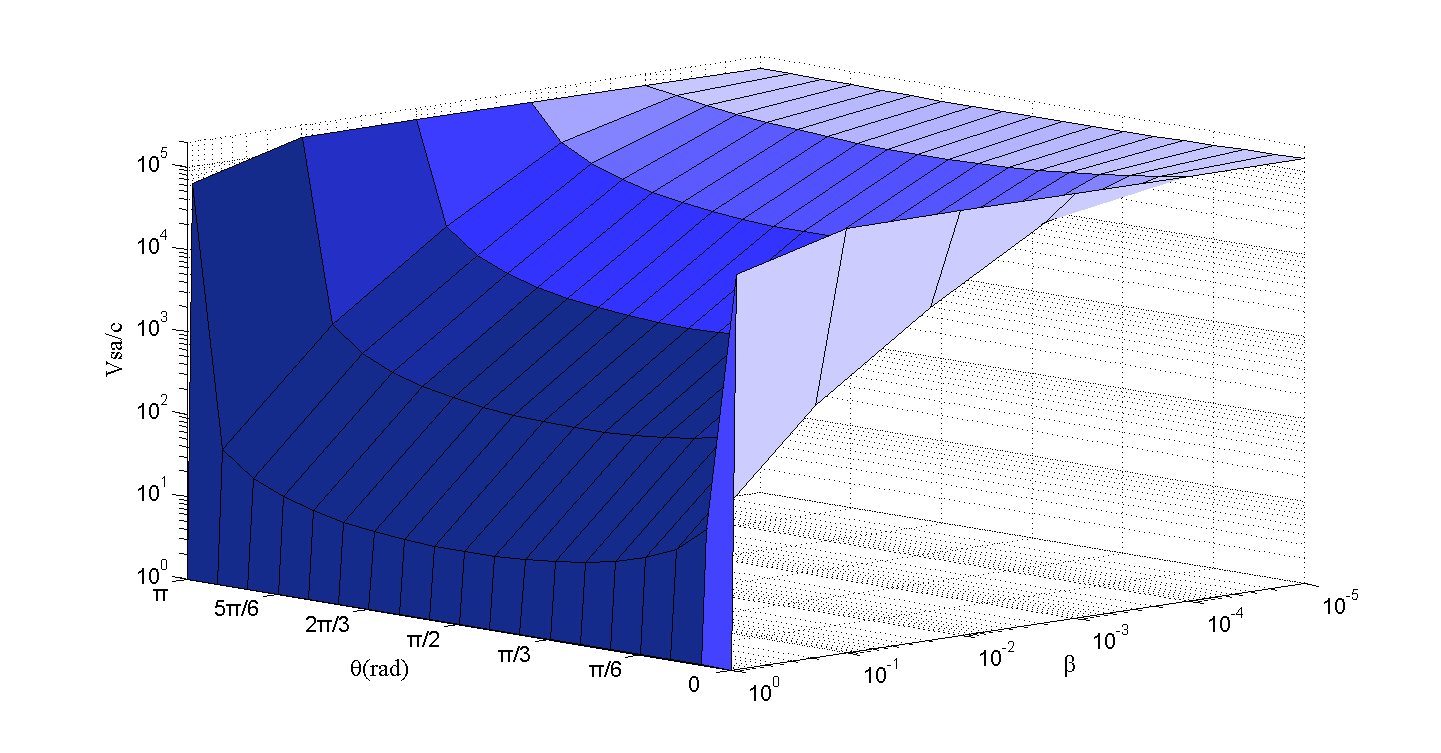}
\end{minipage}
\caption{The speed of spooky action at different $\beta$ and $\theta$. \label{fig:Graph4}}
\end{figure}

\end{document}